\documentclass[letter]{aa} 

\usepackage{graphicx,natbib}
\bibpunct{(}{)}{;}{a}{}{,} 
\usepackage[varg]{txfonts}
\usepackage[colorlinks=true,citecolor=blue]{hyperref}
\usepackage[normalem]{ulem}
\usepackage{threeparttable}
\begin{document} 

   \title{Close-in ice lines and the super-stellar C/O ratio in discs around very low-mass stars}

   \author{Jingyi Mah\inst{1}
          \and
          Bertram Bitsch\inst{1}
          \and
          Ilaria Pascucci\inst{2}
          \and
          Thomas Henning\inst{1}
          }

   \institute{Max-Planck-Institut f\"{u}r Astronomie, K\"{o}nigstuhl 17, 69117 Heidelberg, Germany\\
             \email{mah@mpia.de}
             \and
             Lunar and Planetary Laboratory, Department of Planetary Sciences, The University of Arizona, 1629 East University Boulevard, Tucson, AZ 85721, USA
             }

   \date{Received ; }

 
  \abstract
  {The origin of the elevated C/O ratios in discs around late M dwarfs compared to discs around solar-type stars is not well understood. Here we endeavour to reproduce the observed differences in the disc C/O ratios as a function of stellar mass using a viscosity-driven disc evolution model and study the corresponding atmospheric composition of planets that grow inside the water-ice line in these discs. We carried out simulations using a coupled disc evolution and planet formation code that includes pebble drift and evaporation. We used a chemical partitioning model for the dust composition in the disc midplane. Inside the water-ice line, the disc's C/O ratio initially decreases to sub-stellar due to the inward drift and evaporation of water-ice-rich pebbles before increasing again to super-stellar values due to the inward diffusion of carbon-rich vapour. We show that this process is more efficient for very low-mass stars compared to solar-type stars due to the closer-in ice lines and shorter disc viscous timescales. In high-viscosity discs, the transition from sub-stellar to super-stellar takes place faster due to the fast inward advection of carbon-rich gas. Our results suggest that planets accreting their atmospheres early (when the disc C/O is still sub-stellar) will have low atmospheric C/O ratios, while planets that accrete their atmospheres late (when the disc C/O has become super-stellar) can obtain high C/O ratios. Our model predictions are consistent with observations, under the assumption that all stars have the same metallicity and chemical composition, and that the vertical mixing timescales in the inner disc are much shorter than the radial advection timescales. This further strengthens the case for considering stellar abundances alongside disc evolution in future studies that aim to link planet (atmospheric) composition to disc composition.}


   \titlerunning{Close-in ice lines and the super-stellar C/O ratio in discs around very low-mass stars}
   \authorrunning{J. Mah et al.}
   
   \maketitle
%
\section{Introduction}
More and more sub-Neptunes and super-Earths $(R_{\rm p} < 4~R_{\oplus})$ orbiting M dwarfs $(M_* \lesssim 0.6~M_{\odot})$ are being discovered by the Transiting Exoplanet Survey Satellite (TESS) and the Calar Alto high-Resolution search for M dwarfs with Near-infrared and optical Echelle Spectrograph \citep[CARMENES; e.g.][]{Kostovetal2019,Sabottaetal2021,VanEylenetal2021,Kossakowskietal2023,Ribasetal2023}. As M dwarfs are the most common stars in the Galaxy \citep[e.g.][]{Chabrier2005}, planets around these stars could contribute a substantial dataset for atmospheric characterisation, which will become promising in the era of the \textit{James Webb }Space Telescope (JWST). Information about the atmospheric composition of these planets could then be used to further improve planet formation models, with the long-term goal of putting together a comprehensive picture of the complex process of planet formation. 

To be able to link the observed atmospheric compositions of planets to their formation pathway, it is crucial to have a good understanding of the protoplanetary disc in which they form. While the Atacama Large Millimeter/submillimeter Array (ALMA) has enabled surveys of discs down to late M dwarfs \citep[e.g.][]{Ansdelletal2016,Pascuccietal2016}, infrared observations probing the region inside the water-ice line are limited. Nevertheless, previous observations with {\it Spitzer} have demonstrated that discs around late M dwarfs have a distinct volatile content inside the water-ice line compared to discs around solar-type stars: while the former show strong C$_2$H$_2$ emission, a relatively weaker signal from HCN, and no or extremely weak H$_2$O emission \citep[][]{Pascuccietal2009,Pascuccietal2013}, the latter display an opposite trend \citep[e.g.][]{CarrNajita2011,Salyketal2011}. Due to the short vertical mixing timescale in the inner disc, the composition of the atmosphere that is probed by observations should in principle be similar to the composition of the midplane, where planets assemble \citep[e.g.][]{SemenovWiebe2011}. Results from previous observations thus suggested early on that disc composition changes from oxygen rich (low C/O ratio) to carbon rich (high C/O ratio) as stellar mass decreases. 

Recent observations as part of the MIRI mid-INfrared Disk Survey (MINDS) with the JWST Mid-Infrared Instrument (MIRI) confirm this trend of increasing C/O ratio with decreasing stellar mass. The spectra of inner discs around stars with $M_* \sim 0.75~M_{\odot}$ exhibit strong water features \citep{Gasmanetal2023,Perottietal2023}, while emissions originating from hydrocarbons dominate the inner disc of a $0.14~M_{\odot}$ star \citep{Taboneetal2023}. In the `borderline case', the column densities for H$_2$O and CO$_2$ in the inner disc around a $0.46~M_{\odot}$ star are reported to be similar \citep{Grantetal2023}. Many more JWST/MIRI spectra of discs around M dwarfs are expected to be obtained and analysed in the coming years. 

The origin of the relatively high abundance of carbon-bearing molecules in discs around late M dwarfs compared to discs around solar-type stars is not well understood. Some studies attribute this to the low stellar irradiation intensity of late M dwarfs, which prevents the dissociation of these molecules \citep[e.g.][]{Pascuccietal2009,Walshetal2015}, while others suggest that this could be due to the destruction of carbon grains and polycyclic aromatic hydrocarbons in the inner disc \citep[e.g.][]{Kressetal2010,Andersonetal2017}.

Disc thermo-chemical models that assume an initial solar C/O ratio are able to reproduce the enhanced abundances of C$_2$H$_2$ and HCN but are not able to match the observed column density of C$_2$H$_2$ \citep{Walshetal2015}. To boost the abundance of C$_2$H$_2$ in the disc, one might need to assume that the inner disc’s C/O ratio is high $(\sim 1)$, a notion that is in fact supported by the high ratio of nitriles to water \citep[$\sim 5$;][]{Najitaetal2011}. 

One mechanism that has been suggested to produce a high C/O ratio in the inner region of discs around very low-mass stars is the formation of planetesimals in the outer disc that would trap water ice \citep{CarrNajita2011,Najitaetal2011,Pascuccietal2013}. This explanation would require that the planetesimals predominantly form between the water and CO$_2$ ice lines so that they only lock water ice and not carbon-bearing species. If planetesimals were to also form outside the CO$_2$ ice line, they would trap carbon in them, resulting in no overall change to the disc C/O ratio. However, the process of planetesimal formation near the water-ice line should also take place in discs around higher-mass stars \citep{DrazkowskaAlibert2017}.

Alternatively, the inward drift and evaporation of pebbles at ice lines could also enrich the disc with volatiles and consequently change the disc's C/O ratio. In the outer disc, observations of HD 163296 suggest that the strongly super-stellar C/H ratio is due to this effect \citep{Zhangetal2020}. For the inner disc regions, \citet{Banzattietal2020} suggest that pebble drift and evaporation could also play a role in influencing the volatile content, especially that of water. Furthermore, this effect also seems to be very important for the composition of growing planets \citep[e.g.][]{Boothetal2017,SchneiderBitsch2021a,SchneiderBitsch2021b,Bitschetal2022}.

In this work we show that the observed dependence of the disc C/O ratio on stellar mass could be a direct consequence of disc evolution. To do this, we performed disc evolution and planet formation simulations that take viscous evolution, dust growth, drift, and evaporation into account. We used a chemical partitioning model to compute the initial dust composition. We also discuss what our results imply for the atmospheric composition of planets that form in close-in regions. 

\section{Model}
\label{sec:methods}
We employed the \texttt{chemcomp} code described in \citet{SchneiderBitsch2021a} for our numerical simulations. The code is a 1D viscous $\alpha$-disc model \citep{ShakuraSunyaev1973,Lynden-BellPringle1974} coupled to a planetary growth model. Dust grains and pebbles in the disc grow, drift, and fragment following the two-populations model of \citet{Birnstieletal2012}. 

The dust composition in the disc midplane was computed following a chemical partitioning model \citep{Madhusudhanetal2014,BitschBattistini2020,SchneiderBitsch2021a}, described in Sect.~\ref{appendix:disc_comp}. At the location of ice lines in the disc (see Sect.~\ref{appendix:disc_temp}), volatiles in the pebbles evaporate and enrich the gas \citep{SchneiderBitsch2021a}. The vapour of these volatiles can also diffuse outwards and recondense to make new pebbles, thereby enhancing their abundance in the disc. 

We did not model the detailed chemical reactions (formation or destruction of molecules) that can occur on the dust grains and in the gas disc. This is motivated by recent results that show that the chemical reaction timescale is much longer than the dust's radial drift timescale \citep[e.g.][]{BoothIlee2019,EistrupHenning2022}. 

\subsection{Initial conditions: Disc evolution}
To probe the gas disc carbon-to-oxygen abundance ratio as a function of stellar mass, we set up simulations of star-disc systems with $M_* = 0.1,\,0.3,\,0.5,\,0.7,\,1.0~M_{\odot}$. The corresponding stellar luminosities were obtained from the stellar evolution models of \citet{Baraffeetal2015}. We chose the luminosities at 1~Myr for this work. Assuming higher or lower luminosities would change the overall disc temperature and therefore the locations of the ice lines. This would not affect the inner regions of the disc (where the water-ice line is located) because viscous heating is the dominant heating mechanism. Since all stars are expected to cool in the first megayear \citep{Baraffeetal2015} and since the ice lines in discs around smaller stars are always closer in compared to their counterparts in discs around more massive stars, we expect the general trend of increasing C/O ratio with decreasing stellar mass (presented in the following section) to remain valid. The only difference would be the absolute timescale for the carbon-rich gas to arrive at the inner disc. 

We fixed the disc mass to $M_{\rm disc} = 0.1~M_*$. Here we make the assumption that the star-disc system starts at an earlier stage (Class 0/I), when the disc mass is presumably more massive than what is measured from observations. We also carried out additional simulations with $M_{\rm disc} = 0.01~M_*$ and show that the general trend remains even when we start with a lower value for the initial disc mass (see Sect.~\ref{appendix:less_massive_disc}). The initial characteristic disc radius was set using the scaling relation $R_{\rm c} = 50~{\rm AU}~(M_*/0.1~M_{\odot})^{0.7}$, modified from the original in \citet{Andrewsetal2018scaling}. With this scaling, $R_{\rm c} = 250~{\rm AU}$ for $M_* = 1~M_{\odot}$, similar to the average disc radius measured with ALMA from the main CO isotopologue \citep[e.g.][]{Sanchisetal2021}. We assumed a solar composition for the central star (see Table~\ref{tab:comp_sun}) and a fixed stellar metallicity of $Z = 0.0142$ \citep{Asplundetal2009}. The stellar C/O ratio is 0.55. Our initial conditions are summarised in Table~\ref{tab:initial_condition}.

\begin{table}
\centering
    \caption{Initial conditions of our simulations.}
    \label{tab:initial_condition}
    \begin{tabular}{c c c c c}
    \hline\hline
    $M_*~[M_{\odot}]$ & $L_*~[L_{\odot}]$ & $M_{\rm disc}~[M_{\odot}]$ & $R_{\rm c}~[{\rm AU}]$ & $Z$\\ 
    \hline
    0.1 & 0.067 & 0.01 & 50  & 0.0142\\
    0.3 & 0.332 & 0.03 & 110 & 0.0142\\
    0.5 & 0.663 & 0.05 & 155 & 0.0142\\
    0.7 & 1.093 & 0.07 & 195 & 0.0142\\
    1.0 & 1.933 & 0.10 & 250 & 0.0142\\
    \hline
\end{tabular}
\end{table} 

We assumed that the disc initially has the same composition as the central star. The elements are then partitioned into volatile and refractory molecules, with CO, CH$_4$, CO$_2$, and carbon grains as the major carbon-bearing species in the disc midplane and water as the major oxygen-bearer (see Table~\ref{tab:chempartition} for our chemical partitioning model). We allocated 60\% \citep[the interstellar medium abundance;][]{Berginetal2015} of the total carbon available in the disc into refractory carbon grains. In addition, we also explored the case where the fraction of CH$_4$ in the disc is reduced to about 1\% and the excess carbon is allocated to CO. In this scenario the abundance of water is also reduced because the larger amount of CO takes away more oxygen that is needed to form water.

For each stellar mass, we used a fixed pebble fragmentation velocity of $u_{\rm frag} = 5~{\rm m\,s^{-1}}$ (but also ran additional simulations for 1~${\rm m\,s^{-1}}$ and 10~${\rm m\,s^{-1}}$) and varied the disc viscosity between $\alpha = 10^{-4}$ and $10^{-3}$. All simulations were run for 10~Myr with a time step of 10~yr. We recorded the composition of the disc every 0.05~Myr.

\subsection{Initial conditions: Planet formation}
To investigate the possible differences in the atmospheric compositions of planets that grow inside the ice line in discs around stars of different masses, we implanted planetary cores at $r = 0.3~{\rm AU}$ (inside the water-ice line for all stellar masses investigated in this work) at $t = 0.05~{\rm Myr}$ with initial masses equal to the pebble isolation mass so that the planetary cores start at the gas accretion phase. 

We also assumed a constant gas accretion rate of $\Dot{M}_{\rm gas} = 0.1~M_{\oplus}/{\rm Myr}$ and omitted the effects of orbital migration. We chose this approach for the sake of simplicity because gas accretion in the inner disc region is an active field of study: the accretion rates depend on the core mass and the opacity in the envelope \citep[e.g.][]{Ikomaetal2000,BitschSavvidou2021,Brouwersetal2021} and additionally on recycling flows \citep{Cimermanetal2017,LambrechtsLega2017,Moldenhaueretal2021}.

\section{Results}
\label{sec:results}
\subsection{Gas disc C/O ratio as a function of stellar mass}
\label{sec:results_disc}

\begin{figure*}
\centering
   \resizebox{\hsize}{!}{\includegraphics{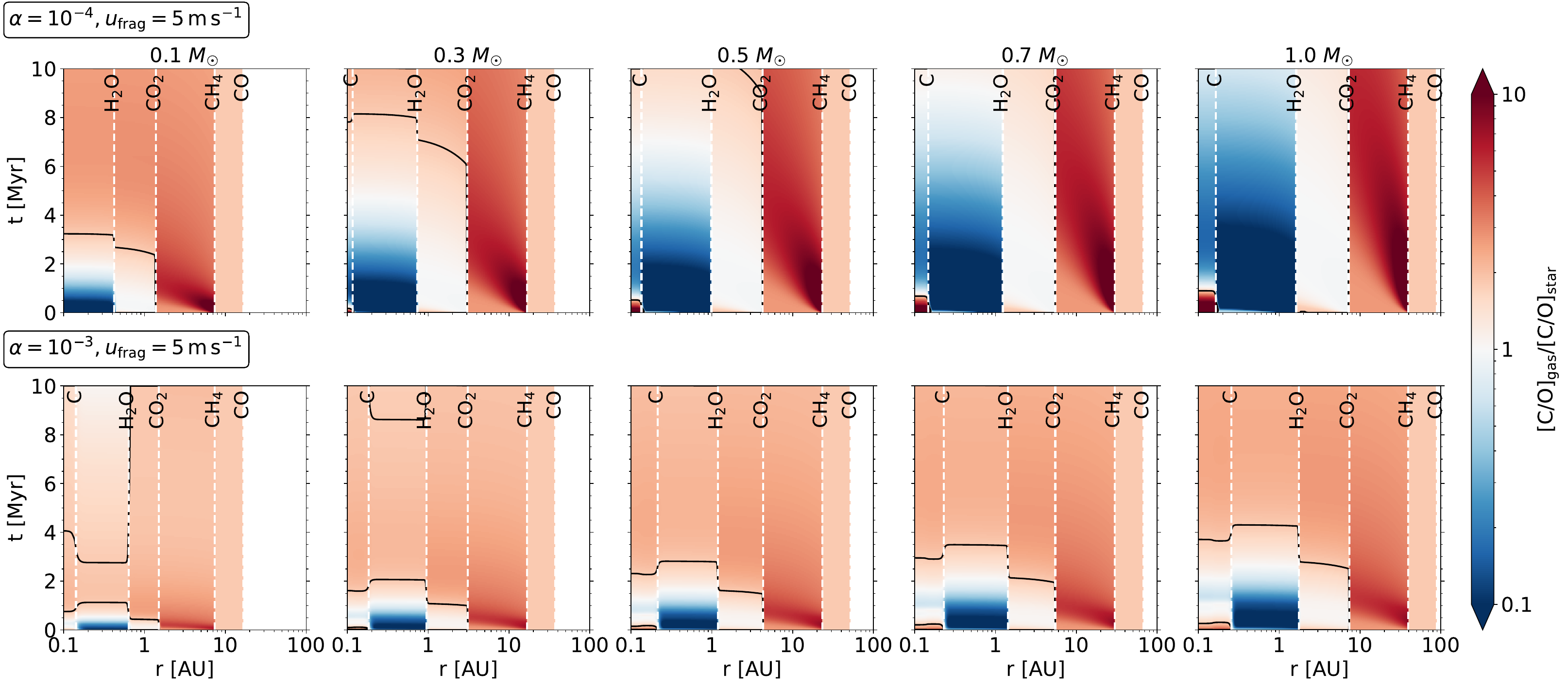}}
   \caption{Temporal evolution of the gas disc C/O ratio (normalised to the stellar value) as a function of distance from the star, stellar mass (left to right), and disc viscosity (top and bottom rows). Ice lines are marked by vertical dashed white lines, and solid black lines indicate an absolute C/O ratio of 1. The carbon line is absent in the top leftmost panel because it is located inside 0.1~AU. Outside the CO ice line ($T_{\rm disc} \leq 20~{\rm K}$; white region in the subplots), all chemical species are assumed to be frozen in the solid phase.}
   \label{fig:discCO_C60_5ms}
\end{figure*}

In Fig.~\ref{fig:discCO_C60_5ms} we show the evolution of the gas disc C/O ratio (normalised to the stellar value) as a function of orbital distance, stellar mass, and disc viscosity. In general, the gas disc C/O ratio ranges from stellar to super-stellar outside the water-ice line ($T_{\rm disc} < 150~{\rm K}$) due to the presence of vapour produced by the evaporation of carbon-bearing species (CO, CH$_4$, and CO$_2$ in our model). In contrast, the ratio is sub-stellar between the water-ice line and the carbon line, where water, the major oxygen-bearer in our model, evaporates. Inside the carbon line $(T_{\rm disc} \geq 631~{\rm K})$, the disc is initially super-stellar in C/O but transitions quickly to sub-stellar due to the inward advection of gas that is sub-stellar in C/O.

More importantly, our simulations show that inside the water-ice line, the time it takes for the C/O ratio to change from sub-stellar to super-stellar becomes shorter as the stellar mass decreases. The trend is clearest when disc viscosity is low ($\alpha = 10^{-4}$): for discs around $M_* = 1~M_{\odot}$, this region can remain sub-stellar in C/O for as long as 10~Myr, whereas the initial sub-stellar C/O signature changes to super-stellar within 2~Myr when the stellar mass is $0.1~M_{\odot}$ (top row of Fig.~\ref{fig:discCO_C60_5ms}). 

Two factors contribute hand-in-hand to the fast transition from sub-stellar to super-stellar C/O in the inner disc of a $0.1~M_{\odot}$ star: the close-in ice lines and the short viscous timescale of the disc. The close-in ice lines are the consequence of a low stellar luminosity and a low mass accretion rate \citep[$\Dot{M}_{\rm acc}$ scales as $\propto M_*^2$; e.g.][]{Manaraetal2017}, which result in weak viscous heating in the disc and therefore a low disc midplane temperature (see Sect.~\ref{appendix:disc_temp}). As a result, super-stellar C/O gas outside the water-ice line only needs to cross a short distance to arrive at the inner disc for very low-mass stars.

The short viscous timescale of discs around $M_* = 0.1~M_{\odot}$ is related to their large scale heights (see Sect.~\ref{appendix:t_visc}). This is because the viscous timescale of a disc is inversely proportional to its kinematic viscosity $(t_{\rm visc} \propto 1/\nu)$ and the disc's kinematic viscosity is in turn proportional to its scale height $(\nu \propto H)$. This results in $t_{\rm visc} \propto 1/H$. As a consequence of the short viscous timescale, sub-stellar C/O gas inside the water-ice line (contributed by water vapour) is removed quickly by accretion onto the central star and is replaced by super-stellar C/O gas (contributed by CH$_4$ vapour), causing the initial sub-stellar signature of the inner disc to be overwritten. In the case of $\alpha = 10^{-4}$, the absolute C/O ratio of the disc inside the water-ice line in our fiducial model reaches 1 at about 3.3 Myr and can subsequently attain an absolute C/O > 1 (top-left panel of Fig.~\ref{fig:discCO_C60_5ms}). The maximum attainable C/O ratio depends on the assumed initial abundance of CH$_4$ in the disc. Vapour of CO and CO$_2$ cannot raise the absolute C/O ratio to above 1 because they have absolute C/O ratios of 1 and 0.5, respectively. We discuss this further in Sect.~\ref{sec:discussion_disc_comp}. 

Our results imply that for discs whose evolution is mainly driven by viscosity, there should be an observable difference in the C/O ratio inside the water-ice line between discs around very low-mass and solar-type stars for nearly all ages. Discs around very low-mass stars are expected to be super-stellar in C/O inside the water-ice line unless we happen to observe them when they are young (< 2~Myr), whereas we expect a sub-stellar C/O ratio inside the water-ice line for discs around solar-type stars. Furthermore, we would also be able to distinguish between discs of low and high viscosities if we knew both the stellar mass and disc age.

\subsection{Atmospheric C/O ratio of close-in planets}
\label{sec:results_planet}

\begin{figure*}
\centering
   \resizebox{\hsize}{!}{\includegraphics{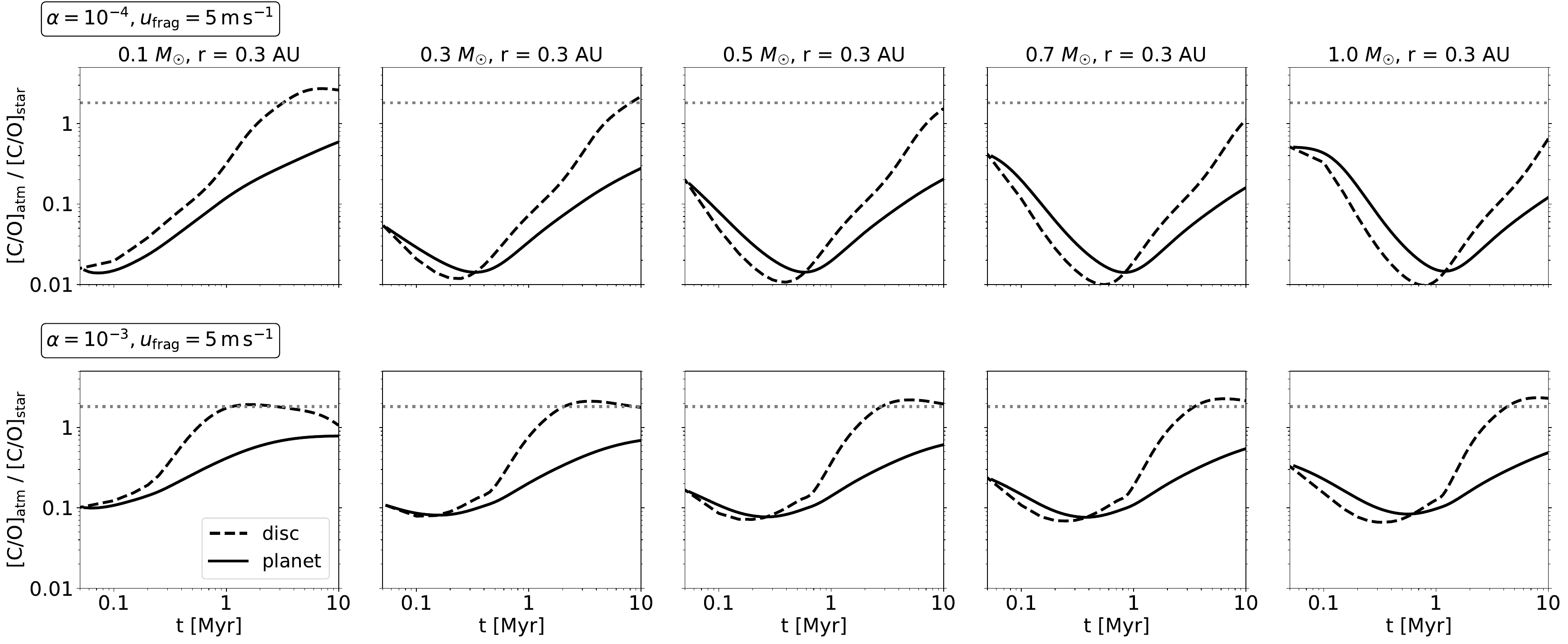}}
   \caption{Temporal evolution of the atmospheric C/O ratio (solid lines) of a non-migrating planet located at $r = 0.3~{\rm AU}$ and the disc C/O ratio (dashed lines) at the same location as a function of stellar mass (left to right) and disc viscosity (top and bottom rows). Ratios are normalised to the stellar value, as in Fig.~\ref{fig:discCO_C60_5ms}. Dotted lines indicate an absolute C/O = 1.}
   \label{fig:planetCO_C60_5ms}
\end{figure*}

In Fig.~\ref{fig:planetCO_C60_5ms} we plot the time evolution of the atmospheric C/O ratio of a planet that forms at 0.3~AU (inside the water-ice line for all stellar masses that we investigated here) as well as the C/O ratio of the disc at the same location for comparison. With the exception of the case for $M_* = 0.1~M_{\odot}$, the C/O ratios of both the planetary atmosphere and the disc first decrease before increasing again. The initial decrease in the C/O ratio is due to the evaporation of water-ice-rich pebbles, and the subsequent increase is caused by the inward advection of carbon-rich gas coupled with the continuous mass loss onto the central star.

Under the assumption of a constant gas accretion rate, we find that the C/O ratio of the planetary atmospheres started with a sub-stellar value and remained sub-stellar until the end of the simulation. At $t = 10~{\rm Myr}$, the atmosphere of a planet around a $0.1~M_{\odot}$ star has a higher C/O ratio compared to that of a planet around a $1~M_{\odot}$ star (0.6 vs 0.12 times the stellar value). When the disc viscosity is high $(\alpha = 10^{-3})$, the difference in the planetary atmospheric C/O ratio becomes smaller (0.8 vs 0.5 times the stellar value), but the absolute value of the ratio increases. 

Our simulations show that the C/O ratio of the planetary atmosphere traces that of the surrounding disc but with a slight time lag. This `lag' of the planetary C/O ratio is determined by our assumption of the gas accretion rate onto the planet. In reality, the gas accretion rate varies in time, and the timing when most of the atmosphere is accreted would exert an influence on the atmospheric C/O ratio: if most of the gas is accreted early when the disc is still sub-stellar in C/O, then the planet will have a sub-stellar C/O atmosphere; if most of the gas is accreted late when the disc C/O has changed to super-stellar, then the planet has a high probability of achieving a super-stellar C/O ratio in its atmosphere. 

This result shows that planets are likely to have a sub-stellar C/O atmosphere if they spend most of their gas accretion phase accreting oxygen-rich gas from inside the water-ice line, even for planets that grow in discs around very low-mass stars. On the other hand, it has been shown that a planet's atmospheric C/O ratio does not provide strong constraints on its formation location \citep[e.g.][]{Turrinietal2021,Bitschetal2022,Molliereetal2022,Fonteetal2023}. A planet that forms outside the water-ice line but migrated inwards to the inner disc very early on can also have a sub-stellar C/O signature that is indistinguishable from one that originates from inside the water-ice line but with lower C/H and O/H ratios. The accretion of carbon at a later time would also alter the planetary C/O ratio. Additional information about the abundance of other chemical species \citep[e.g.][]{Lothringeretal2021,SchneiderBitsch2021b,HandsHelled2022,Pacettietal2022,Chachanetal2023} in the planetary atmosphere would be needed to constrain a planet's formation history. 

\section{Factors influencing the disc C/O ratio}
\label{sec:discussion}
\subsection{Disc composition}
\label{sec:discussion_disc_comp}
By considering the inward drift and evaporation of pebbles, the gas C/O ratio in the region between the CO$_2$ and CH$_4$ ice lines can be enhanced to highly super-stellar values (Fig.~\ref{fig:discCO_C60_5ms}). The inward advection of gas with a high C/O ratio originating from this region allows the disc inside the water-ice line to attain a super-stellar C/O ratio. The abundance of CH$_4$ in the disc then becomes important because it determines how much CH$_4$ vapour can be produced from the pebble evaporation and condensation mechanism and therefore the maximum C/O ratio that the gas inside the water-ice line can reach. In our fiducial model, we allocate 10\% of the total available carbon to CH$_4$ \citep{SchneiderBitsch2021b}, yielding a CH$_4$/H$_2$O abundance ratio of 10\% (see Table~\ref{tab:chempartition}). This fraction is slightly higher than the ratio inferred from observations of interstellar ices and comets \citep[1\% to 5\%; e.g.][]{Gibbetal2004,MummaCharnley2011}.

To find out how the initial abundance of CH$_4$ influences the final gas C/O ratio inside the water-ice line, we carried out additional simulations with only 1\% of the carbon allocated to CH$_4$. In this scenario, the CH$_4$/H$_2$O abundance ratio is 1.1\%, which is on the lower end of the observations \citep[e.g.][]{Gibbetal2004,MummaCharnley2011}. The results show that the initial sub-stellar C/O signature inside the water-ice line is still being overwritten by a super-stellar C/O signature for discs around stars with $M_* \leq 0.3~M_{\odot}$, although the C/O ratio is now lower compared to the fiducial case (Sect.~\ref{appendix:less_methane}). We note that the absolute C/O ratio in this case remains below 1 until much later $(\sim 8.5~{\rm Myr})$. 

As observations show a very carbon-rich-dominated chemistry in the inner discs of very low-mass stars \citep{Pascuccietal2009,Pascuccietal2013,Taboneetal2023}, which requires C/O > 1, this suggests that a certain fraction of CH$_4$ (or other carbon-rich molecules that evaporate in the outer disc) is required in our model to be in line with observations. However, several mechanisms that act in different layers of the disc can alter the abundance of CH$_4$: UV photodissociation in the disc atmosphere; conversion to CO or complex hydrocarbons in the molecular layer; and freeze-out in the midplane \citep[e.g.][]{SemenovWiebe2011,HenningSemenov2013}. At the same time, CH$_4$ can also be produced in the disc via other chemical reactions \citep{SemenovWiebe2011}. Further testing via detailed chemical modelling is needed to determine whether the abundance of CH$_4$ can remain stable for long timescales in the disc needs, which is beyond the scope of this work. 

Since our model is a 1D model, the results can be directly compared to observations under two conditions: first, observations are probing the disc midplane where molecules are shielded from dissociation by UV rays, and second, observations are probing the region where vertical mixing is very efficient, for example the inner disc.

It is clear from the results of our additional simulations that the disc and planet C/O ratios depend on the initial distribution of carbon and oxygen into solids (chemical species present in the disc and their abundances) as well as what we assume for the stellar abundances. The latter is perhaps more important as it governs the initial C/O ratio of the disc. We assumed a solar composition for all stars in this work, but we are aware that there is in fact a spread in the distribution of the C/O ratio (from 0.1 to 1.0) as a function of stellar metallicity for solar-type stars \citep[e.g.][]{BitschBattistini2020}. Detailed stellar abundances should be taken into account in future studies of individual exoplanetary systems. 

\subsection{Pebble size}
As the pebble size influences the amount of vapour that is released to the gas as a function of time, it could affect the disc's C/O profile. We ran additional simulations with pebble fragmentation threshold velocities of $u_{\rm frag} = 1~{\rm m\,s^{-1}}$ (smaller pebbles) and 10 ${\rm m\,s^{-1}}$ (larger pebbles) to investigate the sensitivity of our results to variations in pebble size. 

Increasing the size of pebbles in the disc results in more vapour being released to the gas phase, but the faster drift speed of large pebbles also causes them to be removed more quickly from the disc. For the case of $\alpha = 10^{-3}$, we do see that the region inside the water-ice line is sub-stellar for a shorter period of time compared to Fig.~\ref{fig:discCO_C60_5ms} (the fiducial model) due to the faster removal of pebbles, but the overall result is otherwise not very different (Sect.~\ref{appendix:grain_size}). 

The difference in the disc C/O ratio is more pronounced when we limit the pebble size to a smaller value (Sect.~\ref{appendix:grain_size}). Although smaller pebbles release a smaller amount of vapour when they evaporate, they also couple more strongly to the gas. The resulting outcome is therefore dictated by the disc viscosity: when $\alpha = 10^{-4}$ the change in the C/O ratio takes place more slowly compared to Fig.~\ref{fig:discCO_C60_5ms}; when $\alpha = 10^{-3}$ the difference in the C/O ratio inside and outside the water-ice line is smoothed out because it cannot be sustained by the small amount of water vapour released by the pebbles. 

\subsection{Internal photoevaporation}
In our model we only considered the effects of disc viscosity, and we find that this parameter plays a major role in influencing the C/O ratio. However, it is also known that discs are subjected to stellar irradiation in the form of UV and X-rays, which drive mass loss via internal photoevaporation. Depending on the high-energy stellar photons impinging on the disc, photoevaporation can carve out a gap between $\sim 1$ and 10~AU \citep[e.g.][]{Gortietal2009,Ercolanoetal2021,Picognaetal2021}, which can prevent carbon-rich gas in the outer disc from reaching the inner disc.

Depending on the time it takes for photoevaporation to become efficient, the C/O ratio of the inner disc can be either super-stellar or sub-stellar. The former outcome is expected when photoevaporation carves the gap `late', that is, after super-stellar C/O gas from the outer disc has polluted the inner disc. The C/O ratio of the inner disc will then be `frozen' at the instantaneous value at the time of gap opening. The latter outcome is possible if the gap is carved `early', that is, before the arrival of carbon-rich gas. The longevity of the inner disc's sub-stellar C/O ratio is then determined by its viscous timescale.

\subsection{Gap formation by planets}
If a growing planet opens a gap in the protoplanetary disc, it can block inward drifting pebbles \citep[e.g.][]{Lambrechtsetal2014,Ataieeetal2018,Bitschetal2018,Stammleretal2023} and thus alter the composition of the inner disc depending on its position relative to the ice lines \citep{Bitschetal2021}. In particular, a planet exterior to the water-ice line could block water-ice-rich pebbles, resulting in a less sub-stellar C/O ratio of the inner disc. At the same time, carbon-rich gas can still pass through the giant planet because the process of gas accretion onto a planet is not 100\% efficient \citep[e.g.][]{LubowDAngelo2006,BergezCasalouetal2020}. Consequently, the transition from a sub-stellar C/O to a super-stellar C/O in the inner disc could happen at earlier times compared to a scenario where no giant planets form in the disc. We note that this possibility does not largely affect the conclusions of our simulations related to the disc's C/O ratio, because giant planets are rare around FGK stars \citep[the intrinsic occurrence rate is about $25\%$; e.g.][]{Fernandesetal2019,Fultonetal2021} and even rarer around M dwarfs \citep[e.g.][]{Johnsonetal2010,Rosenthaletal2022}. 

\section{Summary}
We have presented an alternative explanation for the observed differences in the C/O ratio of discs around late M dwarfs and solar-type stars based on results from disc evolution models. The inward advection of carbon-rich gas from outside the water-ice line increases the C/O ratio of the inner disc to a super-stellar value. In our model, the maximum C/O ratio attainable inside the water-ice line depends on the initial abundance of CH$_4$ in the disc, which influences the amount of CH$_4$ produced by the pebble evaporation effect. Assuming that the stars have the same metallicity and composition, and that the vertical mixing timescale in the disc is short, our model predicts super-stellar C/O ratios inside the water-ice line for discs around very low-mass stars if we observe them at $t \geq 2~{\rm Myr}$ and sub-stellar C/O ratios at all times for discs around stars $M_* \geq 0.7~M_{\odot}$ if the disc viscosity is low $(\alpha = 10^{-4})$. In high-viscosity discs $(\alpha = 10^{-3})$, the C/O ratio inside the water-ice line is super-stellar after 0.5~Myr for a very low-mass star and after 3~Myr for a solar-type star. Our model predictions are in line with previous {\it Spitzer} results \citep{Pascuccietal2009,Pascuccietal2013} and recent JWST observations \citep[e.g.][]{Gasmanetal2023,Grantetal2023,Perottietal2023,Taboneetal2023}. Forthcoming observations (or surveys) of discs around stars with masses ranging from $0.1-1~M_{\odot}$ and similar metallicities and compositions would allow us to further test the robustness of our model predictions.

\begin{acknowledgements}
      We thank the reviewer for the constructive comments that helped to improve the clarity of this paper. J.M. and B.B. acknowledge the support of the DFG priority program SPP~1992 ``Exploring the Diversity of Extrasolar Planets'' (BI~1880/3-1). B.B. thanks the European Research Council (ERC Starting Grant 757448-PAMDORA) for their financial support. I.P. acknowledges partial support by the National Aeronautics and Space Administration under Agreement No.~80NSSC21K0593 for the program ``Alien Earths''. T.H. is grateful for the support from the European Research Council (ERC) under the Horizon 2020 Framework Programme via the ERC Advanced Grant Origins 83~24~28.
\end{acknowledgements}

\bibliographystyle{aa} 
\bibliography{main} 

\begin{appendix} 
\section{Supplementary table and figures}
\subsection{Disc composition}
\label{appendix:disc_comp}
We assumed the solar abundances reported in \citet{Asplundetal2009} for the dust composition calculations (Table~\ref{tab:comp_sun}). The chemical species we considered in our model, their respective volume mixing ratios, and their initial abundances (with respect to hydrogen) in the disc are shown in Table~\ref{tab:chempartition}. The partitioning model that we adopt here is based on that of \citet{Madhusudhanetal2014}, which was later expanded by \citet{BitschBattistini2020} and \citet{SchneiderBitsch2021a} to include more molecules. In the fiducial model, the abundance of water ice with respect to hydrogen computed from the partitioning model is $2.67\times10^{-4}$.

\begin{table}
\centering
    \caption{Solar abundances \citep{Asplundetal2009} used in our simulations.}
    \label{tab:comp_sun}
    \begin{tabular}{c c}
    \hline\hline
    Element & Abundance\\ \hline
    He/H & 0.085 \\
    O/H  & $4.90\times10^{-4}$ \\
    C/H  & $2.69\times10^{-4}$ \\
    N/H  & $6.76\times10^{-5}$ \\    
    Mg/H & $3.98\times10^{-5}$ \\
    Si/H & $3.24\times10^{-5}$ \\
    Fe/H & $3.16\times10^{-5}$ \\
    S/H  & $1.32\times10^{-5}$ \\ 
    Al/H & $2.82\times10^{-6}$ \\ 
    Na/H & $1.74\times10^{-6}$ \\ 
    K/H  & $1.07\times10^{-7}$ \\ 
    Ti/H & $8.91\times10^{-8}$ \\ 
    V/H  & $8.59\times10^{-9}$ \\ \hline
\end{tabular}
\end{table}

\begin{table*}
\centering
    \caption{Chemical species included in our model, their respective volume mixing ratios, and their initial abundances with respect to hydrogen.}
    \label{tab:chempartition}
    \begin{threeparttable}
    \begin{tabular}{c c c c}
    \hline\hline
    Species & $T_{\text{cond}}$~(K) & Volume mixing ratio & Initial abundance\\ 
    \hline
    CO              & 20   & 0.20 (0.29) $\times$ C/H & $5.38~(7.80) \times 10^{-5}$ \\
    N$_2$           & 20   & 0.45 $\times$ N/H & $3.04 \times 10^{-5}$ \\
    CH$_4$          & 30   & 0.10 (0.01) $\times$ C/H & $2.69~(0.27) \times 10^{-5}$ \\
    CO$_2$          & 70   & 0.10 $\times$ C/H & $2.69 \times 10^{-5}$ \\
    NH$_3$          & 90   & 0.10  $\times$ N/H & $6.76 \times 10^{-6}$ \\
    H$_2$S          & 150  & S/H  & $1.32 \times 10^{-5}$\\
    H$_2$O          & 150  & O/H - (CO/H + 2 $\times$ CO$_2$/H + 8 $\times$NaAlSi$_3$O$_8$/H + & $2.67~(2.43) \times 10^{-4}$ \\
                    &      & 8 $\times$KAlSi$_3$O$_8$/H + 4 $\times$ Mg$_2$SiO$_4$/H + VO/H + \\
                    &      & 3 $\times$ MgSiO$_3$/H + 3 $\times$ Al$_2$O$_3$/H + TiO/H) \\ 
    C               & 631  & 0.60 $\times$ C/H & $1.61 \times 10^{-4}$ \\
    NaAlSi$_3$O$_8$ & 958  & Na/H & $1.74 \times 10^{-6}$ \\
    KAlSi$_3$O$_8$  & 1006 & K/H & $1.07 \times 10^{-7}$ \\
    Mg$_2$SiO$_4$   & 1354 & Mg/H - (Si/H - 3 $\times$ K/H - 3 $\times$ Na/H) & $1.29 \times 10^{-5}$\\
    Fe              & 1357 & Fe/H & $3.16 \times 10^{-5}$ \\
    VO              & 1423 & V/H & $8.59 \times 10^{-9}$ \\
    MgSiO$_3$       & 1500 & Mg/H - 2 $\times$ (Mg/H - (Si/H - 3 $\times$ K/H - 3 $\times$ Na/H)) & $1.39 \times 10^{-5}$ \\
    Al$_2$O$_3$     & 1653 & 0.5 $\times$ (Al/H - (K/H - Na/H)) & $2.23 \times 10^{-6}$ \\
    TiO             & 2000 & Ti/H & $8.91 \times 10^{-8}$\\
    \hline
    \end{tabular}
    \begin{tablenotes}
        \item Volume mixing ratios of each species are based on the works of \citet{Madhusudhanetal2014}, \citet{BitschBattistini2020}, and \citet{SchneiderBitsch2021a,SchneiderBitsch2021b}, and their condensation temperatures are from \citet{Lodders2003}. Solar elemental abundances \citep{Asplundetal2009} are assumed. Values in parentheses are the ratios we used for the additional simulations with less CH$_4$ and the corresponding initial abundance of CO, CH$_4$, and water.
    \end{tablenotes}
    \end{threeparttable}
\end{table*} 

\subsection{Disc temperature profile}
\label{appendix:disc_temp}
\begin{figure*}
\centering
   \resizebox{\hsize}{!}{\includegraphics{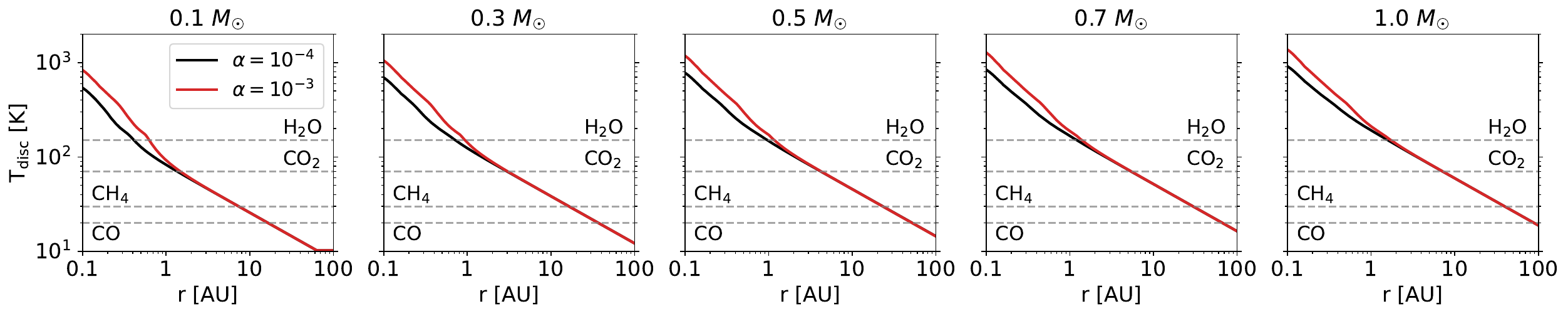}}
   \caption{Disc midplane temperature profile as a function of stellar mass and disc viscosity. The condensation temperatures of carbon- and oxygen-bearing species in the disc are marked with dashed grey lines.}
   \label{fig:disc_temp}
\end{figure*}

We show in Fig.~\ref{fig:disc_temp} the temperature profile of the disc midplane as a function of stellar mass and disc viscosity. As the disc heating sources we considered in our model are viscous heating and stellar irradiation, disc temperatures increase with disc viscosity (in the region inside $\sim 1~{\rm AU}$) as well as stellar mass (and luminosity). We chose the stellar luminosity at 1~Myr from the stellar evolution models of \citet{Baraffeetal2015}. Choosing different values of luminosities would result in a different disc temperature profile and would change the location of the ice lines. We can already see from Fig.~\ref{fig:disc_temp} that the ice lines in an M dwarf disc are closer to the central star but are spread out over a larger distance in a disc around a solar-type star. 

\subsection{Initial disc mass}
\label{appendix:less_massive_disc}
\begin{figure*}
\centering
   \resizebox{\hsize}{!}{\includegraphics{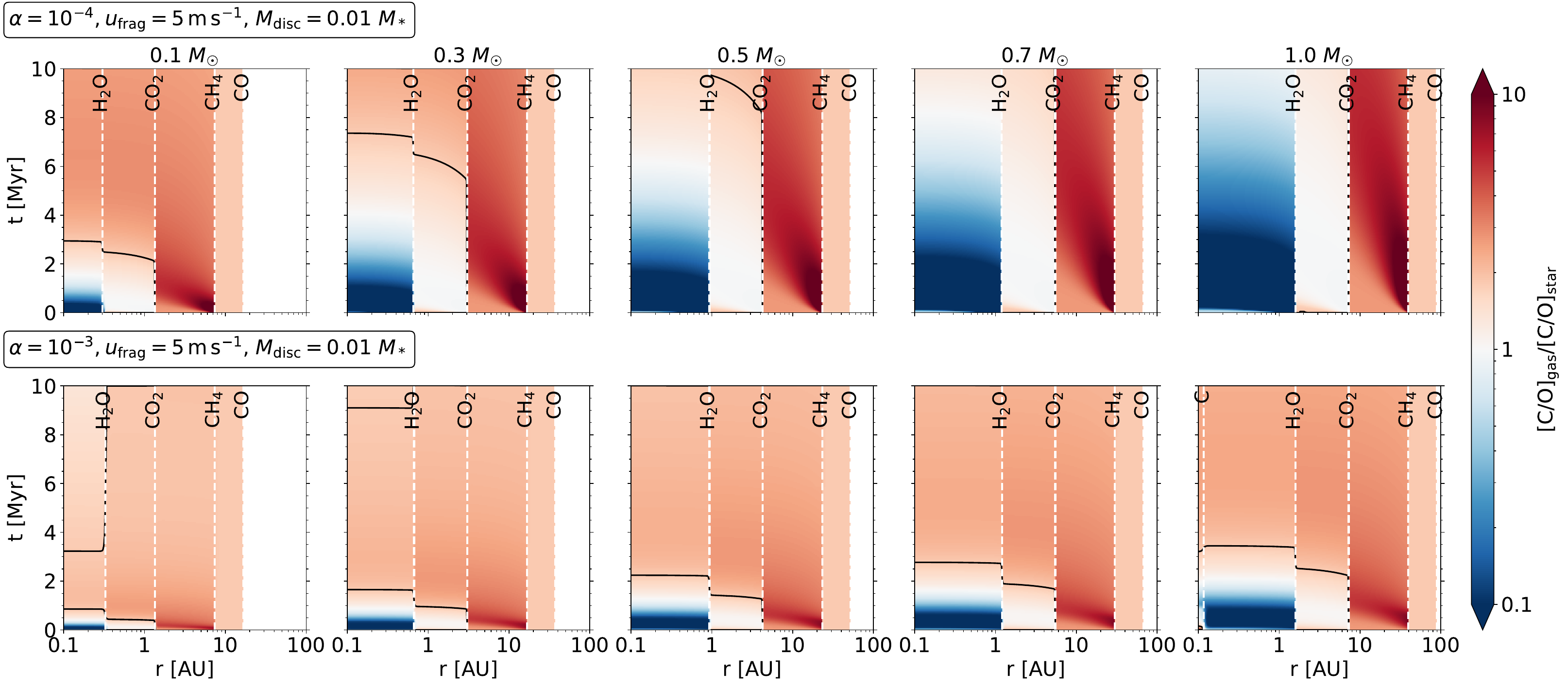}}
   \caption{Same as Fig.~\ref{fig:discCO_C60_5ms} but with $M_{\rm disc} = 0.01~M_*$.}
   \label{fig:discCO_C60_001}
\end{figure*}

Figure~\ref{fig:discCO_C60_001} shows the time evolution of the disc C/O ratio when the initial disc mass is assumed to be 1\% of the stellar mass. When the disc mass is reduced, the gas surface density and therefore the heat due to viscous heating also decrease. Since part of the disc midplane temperature is contributed by viscous heating, the disc temperature in the viscous heating region $(r \lesssim 2~{\rm AU})$ becomes lower when we reduce the initial disc mass. Consequently, the location of ice lines inside $\sim 2~{\rm AU}$ are shifted inwards. In this set of simulations, the carbon line is absent (except for the case of $M_* = 1.0~M_{\odot}$ and $\alpha = 10^{-3}$) because the disc temperature at 0.1~AU is lower than the carbon evaporation temperature. As a result of the closer-in ice lines, carbon-rich gas `pollutes' the region inside the water-ice line faster. The overall trend with stellar mass nevertheless remains.

\subsection{Disc viscous timescale}
\label{appendix:t_visc}
We compare the viscous accretion timescale of a disc around a $0.1~M_{\odot}$ star and a $1.0~M_{\odot}$ star to understand how it influences our results in the main paper. The timescale of viscous accretion is given as
\begin{equation}
    \label{eq:t_visc}
    t_{\rm visc} = \frac{r^2}{\nu} = \frac{r^2}{\alpha H^2 \sqrt{GM_*/r^3}},
\end{equation}
where $r$ is the radial distance from the central star, $\nu$ is the kinematic viscosity, $\alpha$ is the turbulent viscosity parameter \citep{ShakuraSunyaev1973}, $H$ is the scale height of the disc, $G$ is the gravitational constant, and $M_*$ is the stellar mass. Under the assumption that a disc around an M dwarf star and a disc around a solar-type star have the same viscosity and disc scale height (or the aspect ratio $H/r$), the viscous timescale of the M dwarf disc would be roughly 3.16 times longer than that of the solar-type star's disc. However, we assume here, for the sake of consistency, an increasing disc radius with increasing stellar mass. This means that the disc scale height also changes with stellar mass. 

\begin{figure*}
\centering
   \resizebox{0.8\hsize}{!}{\includegraphics{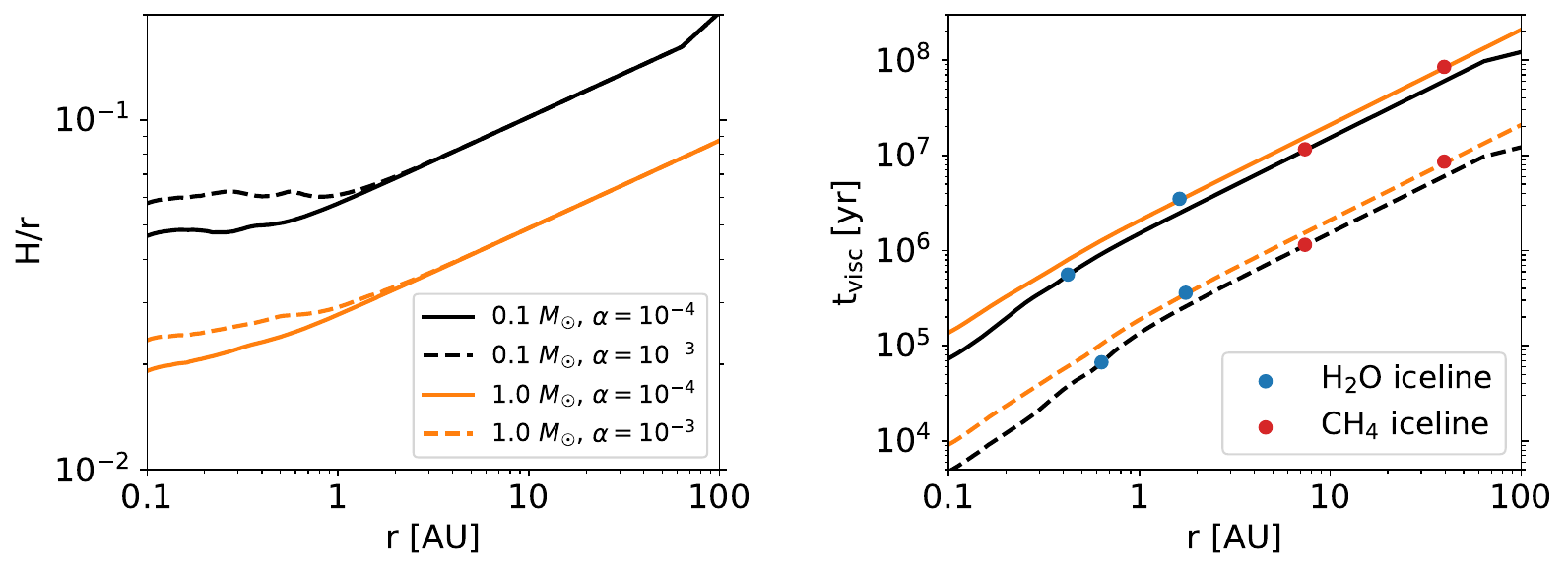}}
   \caption{Comparison of disc properties for a disc around an M dwarf and a solar-type star in our model. {\it Left}: Aspect ratio of a disc around a $0.1~M_{\odot}$ star (black lines) and a disc around a $1.0~M_{\odot}$ star (orange lines) for low (solid lines) and high (dashed lines) disc viscosities. {\it Right}: Viscous accretion timescale of the discs as a function of radial distance, stellar mass, and disc viscosity. Solid circles indicate the location of the water-ice (blue) and CH$_4$-ice (red) lines in each disc.}
   \label{fig:disc_tvisc}
\end{figure*}

In Fig.~\ref{fig:disc_tvisc} we show the aspect ratio of discs around $0.1~M_{\odot}$ and $1.0~M_{\odot}$ and their viscous timescales for comparison. For the same disc viscosity, the disc around a $0.1~M_{\odot}$ star has a higher aspect ratio compared to the disc around a $1.0~M_{\odot}$ star. The higher aspect ratio acts to reduce the viscous timescale of the disc (see Eq.~\ref{eq:t_visc}). On the other hand, the lower stellar mass acts to increase the viscous timescale. The outcome of the competition between the two aforementioned factors shows that for the same disc viscosity, the aspect ratio (or the scale height) exerts a stronger influence than the stellar mass on the viscous timescale (right panel of Fig~\ref{fig:disc_tvisc}). Consequently, the disc around a $0.1~M_{\odot}$ star has a shorter viscous accretion timescale than that of the disc around a $1.0~M_{\odot}$ star in our model $(t_{\rm visc,0.1~M_{\odot}}\approx 0.5\,t_{\rm visc,1.0~M_{\odot}})$.

In addition to the short viscous timescale, the close-in ice lines in an M dwarf disc also contribute to the swift transition from a sub-stellar to a super-stellar C/O ratio in the inner disc. In Fig.~\ref{fig:disc_tvisc}, the corresponding viscous timescales at the location of the ice lines (marked with red and blue circles in the plot) are always shorter by roughly a factor of 10 in the disc around a $0.1~M_{\odot}$ star than in the disc around a $1.0~M_{\odot}$ star.

\subsection{Abundance of methane in the disc}
\label{appendix:less_methane}
\begin{figure*}
\centering
   \resizebox{\hsize}{!}{\includegraphics{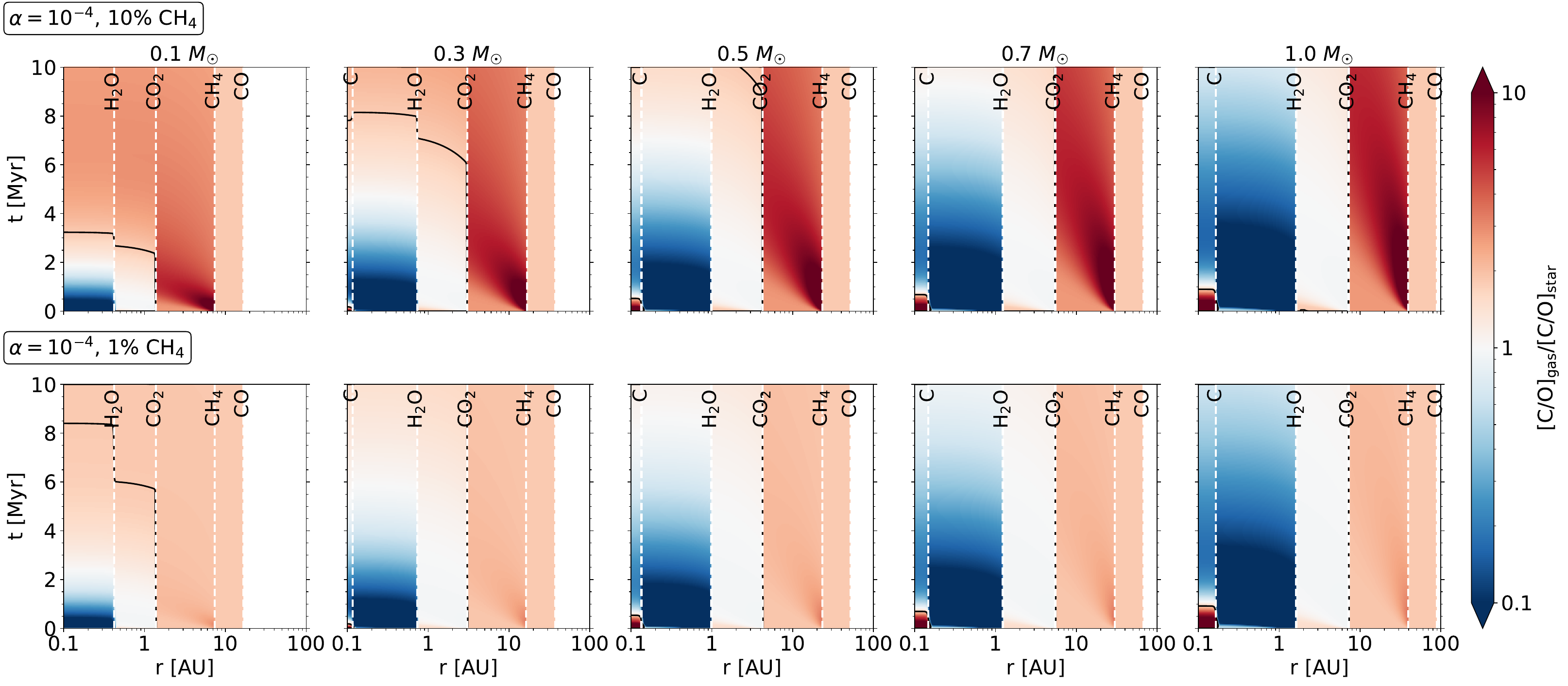}}
   \caption{Comparison of the disc C/O ratio between the fiducial case with 10\% of the total available carbon partitioned into CH$_4$ (top row) and the case with 1\% (bottom row).}
   \label{fig:discCO_C60_lessCH4}
\end{figure*}

Here we report the results of simulations with less CH$_4$ in the disc. We determined this by allocating only 1\% of the available carbon to CH$_4$, with the remainder partitioned to CO (see Table~\ref{tab:chempartition}). The CH$_4$/H$_2$O abundance ratio in this scenario is 0.01. Furthermore, the water abundance is reduced because we now have more CO in the disc, which takes away oxygen needed to form water. We find that the region inside the water-ice line still becomes super-stellar in C/O for discs around stars with mass $M_* \leq 0.3~M_{\odot}$ (Fig.~\ref{fig:discCO_C60_lessCH4}), but the absolute value of the C/O ratio is not as high as for the case when there is more CH$_4$ in the disc, suggesting that the abundance of carbon-rich molecules (in our model we only consider CH$_4$) plays a role in determining the maximum C/O ratio that the inner disc can reach.

\subsection{Grain size}
\label{appendix:grain_size}
\begin{figure*}
\centering
   \resizebox{\hsize}{!}{\includegraphics{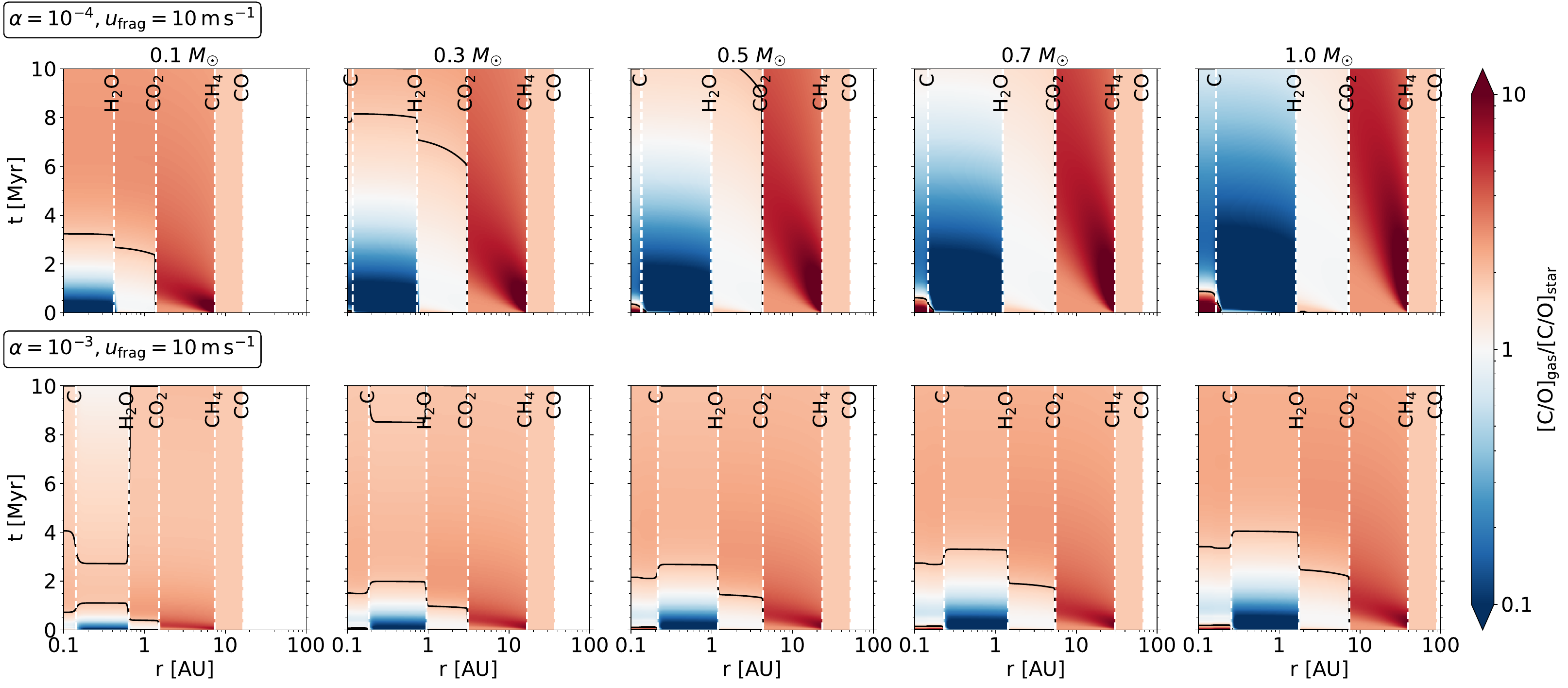}}
   \caption{Same as Fig.~\ref{fig:discCO_C60_5ms} but with the pebble fragmentation velocity increased to 10~${\rm m\,s^{-1}}$.}
   \label{fig:discCO_C60_10ms}
\end{figure*}

We studied the effect of different grain sizes by varying the pebble fragmentation velocity in the code. A larger grain size results in a faster drift of pebbles towards the star. In Fig.~\ref{fig:discCO_C60_10ms} we see that the time when the gas inside the water-ice line is sub-stellar in C/O (blue regions in the figure) is slightly shorter compared to the fiducial model, but the overall trend with stellar mass and disc viscosity does not change.

\begin{figure*}
\centering
   \resizebox{\hsize}{!}{\includegraphics{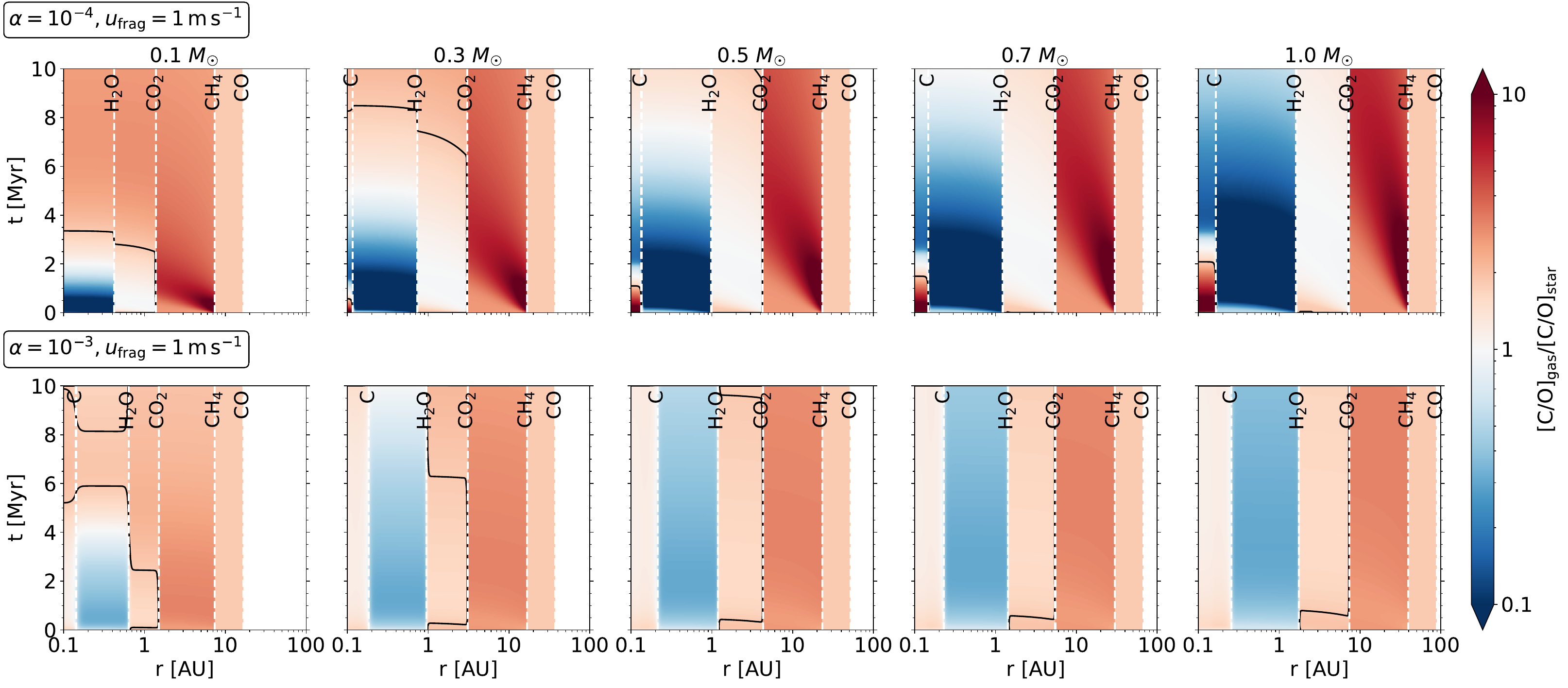}}
   \caption{Same as Fig.~\ref{fig:discCO_C60_5ms} but with the pebble fragmentation velocity reduced to 1~${\rm m\,s^{-1}}$.}
   \label{fig:discCO_C60_1ms}
\end{figure*}

If the disc contains very small grains, then we expect an opposite trend, where the inner disc remains sub-stellar for a longer period of time because the grains are more strongly coupled to the gas. Indeed, this is what our simulation results show (Fig.~\ref{fig:discCO_C60_1ms}).

\end{appendix}
\end{document}